\documentclass[aps,preprint,amsmath,amssymb]{revtex4}
\usepackage{graphicx,color}
\usepackage{subfigure}
\def\be{\begin{eqnarray}}
\def\ed{\end{eqnarray}}
\def\non{\nonumber}

\begin{document}

\title{$SU(2)_X $ Vector DM  and Galactic Center Gamma-Ray Excess   }

\author{ \bf Chuan-Hung Chen$^{a}$\footnote{Email:
physchen@mail.ncku.edu.tw} and Takaaki Nomura$^{a}$\footnote{Email: nomura@mail.ncku.edu.tw} }

\affiliation{ $^{a}$Department of Physics, National Cheng-Kung
University, Tainan 701, Taiwan  }

\date{\today}

\begin{abstract}

An unbroken $Z_3$ symmetry remains when a local $SU(2)_X$ symmetry
is broken spontaneously by a quadruplet. The gauge boson $\chi_\mu (\bar \chi_\mu )$ carries the dark charge  and is the candidate of dark matter (DM).  Due to the mixture of the scalar boson $\phi_r$ of the quadruplet and the standard model (SM) Higgs boson,  the DM can annihilate into SM particles through the Higgs portal. To investigate the implications of the vector DM in the model, we study the relic density of DM, the direct detection of the DM-nucleon scattering and the excess of the gamma-ray spectrum from the Galactic Center, which is supported by the data from the {\it Fermi} Gamma-Ray Space Telescope. We find that with the DM mass of around $70$ GeV in our model, the data for the excess of the gamma-ray could be fitted well.   

\end{abstract}

\maketitle

One of unsolved problems  in  astrophysics is  the existence of  dark matter (DM), where  the plausible candidates in particle physics are the weakly interacting massive particles (WIMPs).  
The Planck best-fit  for  the DM  density,  which combines the data of the
WMAP polarization at low multipoles, high-$\ell$ experiments and baryon acoustic oscillations (BAO),  
 is given by 
\cite{Ade:2013ktc}
 \be
 \Omega_{\rm DM} h^2=0.1187\pm0.0017\,. \label{eq:omega}
 \ed

Besides the evidence from astronomical observations, now there are direct and indirect ways to detect  DM. 
According to the recent measurements by  XENON100~\cite{XENON100} and  LUX~\cite{Akerib:2013tjd} Collaborations, which are designed for directly detecting DM,  since no clear signal is found, the cross section for the elastic scattering of DM  off  nucleons  has been strictly limited.   
Additionally, although the potential DM signals  are indicated  by the indirect detections, such as the excess of the positron fraction observed by 
 PAMELA~\cite{Adriani:2008zr} and Fermi-LAT~\cite{FermiLAT:2011ab} experiments, and the excess of the positron+electron flux observed by PAMELA~\cite{Adriani:2011xv}, Fermi-LAT~\cite{Ackermann:2010ij}, 
ATIC~\cite{Chang:2008aa}, and HESS~\cite{Aharonian:2008aa, Aharonian:2009ah},  they may also be solved by astronomical effects, e.g. pulsars \cite{pulsar1,pulsar2}. 

Recently, a clear excess of the gamma-ray spectrum, which has an obvious peak at the photon energy of around 2 GeV, has been pointed out by  the analyses in Refs.~\cite{Goodenough:2009gk,Hooper:2010mq,Boyarsky:2010dr,Hooper:2011ti,Abazajian:2012pn,Gordon:2013vta,Macias:2013vya,Abazajian:2014fta}. Furthermore, using the data from the observation of  the {\it Fermi} Gamma-Ray Space Telescope~\cite{Vitale:2009hr, Morselli:2010ty}, a more  significant signal of the gamma-ray from the region around the Galactic Center is also found~\cite{Daylan:2014rsa,Zhou:2014lva,Calore:2014xka,Abazajian:2014hsa,Calore:2014nla}.
 Subsequently, it has been  found that  the excess  matches well  with the gamma-ray spectrum  from the DM annihilation, where the requested thermally averaged cross section $\langle \sigma v_{\rm rel} \rangle$  at the order of $10^{-26}$ cm$^3$/s  is the same as that of the  thermal relic density. Moreover, it has been pointed out that the effects through the Higgs portal could naturally explain the excess of the gamma-ray spectrum~\cite{Daylan:2014rsa,Berlin:2014pya}. 
 Based on these results, in this paper we propose a stable vector DM model in which a discrete symmetry stabilizing the DM is obtained naturally and the DMs annihilate into SM fermions through the Higgs portal.
Other mechanisms  to explain the excess could be referred to the references in the literature, such as that DM annihilates directly into SM particles and/or DM first annihilates into hidden scalar (gauge) bosons, and then decays to SM particles via the Higgs ($Z'$)-portal~\cite{Modak:2013jya, Boehm:2014hva, Berlin:2014pya, Boehm:2014bia, Ko:2014gha, Abdullah:2014lla, Ghosh:2014pwa,Martin:2014sxa, Berlin:2014tja,Cline:2014dwa,Wang:2014elb,Arina:2014yna,Cheung:2014lqa,Huang:2014cla,Balazs:2014jla,Baek:2014kna,Okada:2014usa,Ghorbani:2014qpa,Banik:2014eda,Borah:2014ska,Cahill-Rowley:2014ora,Guo:2014gra,Cao:2014efa,Cheung:2014tha,Dolan:2014ska,Biswas:2014hoa, Ghorbani:2014gka,Cerdeno:2015ega}. 

 From the view point of model buildings, to protect DM from its decay, an unbroken symmetry in the theory is necessary. However, a discrete symmetry usually is put in by hand. In order to get a stable DM naturally, we study the model in which the unbroken symmetry originates from a spontaneously broken gauge symmetry. To realize the concept, particularly we are interested in the extension of the SM with a new $SU(2)_X$ gauge symmetry where the subscript $X$ is regarded as a dark charge.  The interesting properties of a local $SU(2)_X$ group are: (1) comparing with the local $U(1)$ case in which the $U(1)$ charge has to satisfy some artificial tuning~\cite{Krauss:1988zc},   an unbroken  discrete symmetry can be naturally preserved  after the spontaneous breaking of the $SU(2)_X$ gauge symmetry;
  (2) the massive gauge bosons from $SU(2)_X$ could be the DM candidates. The various applications of the hidden $SU(2)$ gauge symmetry have been studied in the literature, 
such as  a remaining $Z_2$ symmetry with a quintet in Ref.~\cite{Chiang:2013kqa}, a custodial symmetry in Refs.~\cite{Boehm:2014bia, Hambye:2008bq} and an unbroken $U(1)$ of $SU(2)$ in Refs.~\cite{Khoze:2014woa, Baek:2013dwa}.
 
 Since  the model with the custodial symmetry discussed in Ref.~\cite{Hambye:2008bq} is similar to our proposal, it is worthy 
 to show  the difference between them. It has been noticed that without introducing any new fermions or higher multiple states in  the hidden $SU(2)_X$ gauge sector,  a new fundamental representation of $SU(2)_X$ could lead to three degenerate DM candidates by utilizing the $SO(3)$ custodial symmetry~\cite{Hambye:2008bq}.  Due to the custodial symmetry, the three DM candidates are stable particles. However, the symmetry could be broken easily when $SU(2)_X$ fermions and/or higher representation scalar fields are included. Although the  inclusion of the new fermionic and/or higher multiple staff is not necessary, if one connects the origin of neutrino masses with the dark sector, the inclusion of the new staff becomes a relevant issue. In order to get over the possible unstable effects when  more phenomenological problems in particle physics are involved, we propose to use a discrete symmetry to stabilize  DM, where the discrete symmetry is not broken by  higher multiplet fields or fermions under $SU(2)_X$. 
Additionally, 
the processes for explaining the gamma-ray excess in our model are different from those dictated by the custodial symmetry~\cite{Boehm:2014bia,Hambye:2008bq}. We will see the differences in the analysis below.   
 Moreover, we find that an  $Z_3$ discrete symmetry indeed remains when $SU(2)_X$ is broken by a scalar quadruplet.
 Based on the introduced quadruplet, we  summarize the characteristics of our model as follows: (a) the unbroken  $Z_3$ symmetry is the remnant of $SU(2)_X$, (b) two gauge bosons $\chi_\mu$ and $\bar\chi_\mu$ carry  the $Z_3$ charge and are the candidates of DM, (c) besides the SM Higgs ($\phi$), only one new scalar boson ($\phi_r$) is introduced, and (d) due to the mixture of  $\phi_r$ and $\phi$, the DM annihilation is through the Higgs portal.
%
  
 In the following, we briefly introduce the model and discuss the relevant interactions with the candidates of DM. 
 To study the minimal extension of the SM that  includes the staff of DM, besides the SM particles and their dictated gauge symmetry, we consider a new local $SU(2)_X$ gauge symmetry and add one quadruplet of $SU(2)_X$ to the model. The introduced quadruplet is not only responsible for the breaking of the new gauge symmetry, but also plays an important role on the communication between dark  and visible sectors. Thus,  the Lagrangian in $SU(2)_X \times SU(2)_L \times U(1)_Y$ is written as
 \be
 {\cal L}&=& {\cal L}_{SM} + \left( D_\mu \Phi_4\right)^\dagger D^\mu \Phi_4 - V(H,\Phi_4) - \frac{1}{4} X^a_{\mu\nu} X^{a \mu \nu} \label{eq:lang}
 \ed
 with 
 \be
 V(H, \Phi_4) &=& \mu^2 H^\dagger H + \lambda (H^\dagger H)^2 + \mu^2_\Phi \Phi^\dagger_4 \Phi_4 + \lambda_\Phi (\Phi^\dagger_4 \Phi_4)^2 + \lambda'  \Phi^\dagger_4 \Phi_4 H^\dagger H \,, \label{eq:vphi4}
  \ed
where ${\cal L}_{SM}$ is the Lagrangian of the SM, $H^T=(G^+, (v+\phi + i G^0)/\sqrt{2})$ is the SM Higgs doublet, $\Phi^T_4= (\phi_{3/2}, \phi_{1/2}, -\phi_{-1/2}, \phi_{-3/2})/\sqrt{2}$ is the quadruplet of $SU(2)_X$, the index $i$ of $\phi_i$ stands for the eigenvalue of the third generator of $SU(2)_X$,  $\phi_{-i} = \phi^*_{i}$, the covariant derivative of $\Phi_4$ is $D_\mu  = \partial_\mu +i g_X T^a X^a_\mu $ with the representations of $T^a$ in the quadraplet, given by
 \be
 T^1 = \frac{1}{2}  \left( \begin{array}{cccc}
   0 & \sqrt{3} & 0 & 0  \\ 
    \sqrt{3} & 0 & 2 & 0  \\ 
    0 & 2 & 0 & \sqrt{3}  \\ 
    0 & 0 & \sqrt{3} & 0 \\ 
  \end{array}  \right)\,, \ \ \ T^2 = \frac{i}{2}  \left( \begin{array}{cccc}
   0 & -\sqrt{3} & 0 & 0  \\ 
    \sqrt{3} & 0 & -2 & 0 \\ 
    0 & 2 & 0 & -\sqrt{3} \\ 
    0 & 0 & \sqrt{3} & 0  \\ 
  \end{array}  \right)\,,
 \ed
and $T^3 = {\rm diag}(3/2, 1/2,  -1/2, -3/2)$, and  the field strength tensor of $SU(2)_X$ is read by $X^a_{\mu \nu} = \partial_\mu X^a_\nu -\partial_\nu X^a_\mu - g_X (\vec X_\mu \times \vec X_\nu)^a$. 

To break  $SU(2)_X$ but preserve a discrete symmetry,   the non-vanishing vacuum expectation value (VEV)  and the associated fields fluctuated 
around the VEV are set to be
 \be
  \langle \phi_{\pm 3/2} \rangle = \frac{v_4 }{\sqrt{2} }\,, \quad 
  \phi_{\pm 3/2} = \frac{1 }{\sqrt{2} }(v_4 + \phi_r \pm i \xi)\,. \label{eq:vev}
  \ed
 When we regard the quadruplet  as the fluctuations from the vacuum $\Phi_0  = ( v_4, 0, 0, v_4)/2$,  
  $\Phi_4$  can be parametrized by using the form
 \be
 \Phi_4 &=& e^{i T^a \alpha^a(x) /v_4 } \bar \Phi_4\,,  \label{eq:Nphi4}\\
 \bar \Phi^T_4 &=& \frac{1}{\sqrt{2}}\left( \bar \phi_r, 0,  0, \bar\phi_r \right)\non
 \ed
 with $\bar \phi_ r= (v_4 + \phi_r)/\sqrt{2}$. In terms of scalar fields $\alpha^a(x)$,
the components of $\Phi_4$ could be  expressed 
  as $\phi_{1/2} =  \sqrt{3} (-\alpha^2(x) + i \alpha^1(x) )/2\sqrt{2}$, $\phi_{-1/2} = \phi^*_{1/2}$ and $\xi= 3/2 \alpha^3(x)$, where we have taken the leading terms in the field expansions.  Eq.~(\ref{eq:Nphi4})  indeed is nothing but a local gauge transformation. Therefore, $\phi_{\pm 1/2}$ and $\xi$ could be rotated away from the kinetic term of $\Phi_4$ and the scalar potential; and they are the unphysical Nambu-Goldstone (NG) bosons  of the 
  local $SU(2)_X$ symmetry breaking. Consequently, we can just employ $\bar\Phi_4$ for exploring the mass spectra of new particles.

 With the breaking pattern  in Eq.~(\ref{eq:vev}), one can find that an $Z_3$ symmetry $U_3 \equiv e^{i  T^3 4\pi/3} = {\rm diag}(1, e^{i2\pi/3}, e^{-2i\pi/3}, 1)$ is preserved by the ground state $\Phi_0$. Under the $Z_3$ transformation, the scalar fields of the quadruplet are transformed as
 \be
\phi_{\pm 3/2} & \longrightarrow&\phi _{\pm 3/2}\,,   \non \\
\phi_{\pm 1/2} & \longrightarrow & e^{\pm i 2\pi/3} \phi_{\pm 1/2}\, . 
  \ed
 That is, $\phi_{\pm 3/2}$ are $Z_3$ blind while $\phi_{\pm 1}$ carry the charges of $Z_3$. To understand the transformations of gauge fields, one can use 
   \be
   T^a X^{\prime a} _\mu = U_3 T^b X^b_\mu U^\dagger_3\,.
   \ed 
In terms of physical states of gauge fields, one can  write 
 \be
T^a X^a_\mu = \frac{1}{\sqrt{2}}(T^+ \chi_\mu + T^- \bar\chi_\mu ) + T^3 X^3_\mu \label{eq:T+-}
\ed
with $T^\pm = T^1 \pm i T^2$ and $\chi_\mu (\bar \chi_\mu)= (X^1_\mu \mp i X^2_\mu)/\sqrt{2}$ where $\bar\chi_\mu$ is regarded as the antiparticle of $\chi_\mu$.
  Using the identity $U_3 T^\pm U^\dagger_3 = \exp( \pm i 4\pi/3 )T^\pm$, the transformations of $\chi_\mu (\bar \chi_\mu)$ and $X^3_\mu$ under $Z_3$ are given by
   \be
 &&  X^3_\mu \longrightarrow X^3_\mu \,, \non \\
  && \chi_\mu ( \bar\chi_\mu) \longrightarrow  e^{\pm i 4\pi/3} \chi_\mu (\bar\chi_\mu)\,.
   \ed
  We see that $\chi_\mu (\bar \chi_\mu)$  carries the $Z_3$ charge and $X^3_\mu$ is the $Z_3$ blind. Due to the unbroken $Z_3$,
   the particles with the charges of $Z_3$ are the candidates of DM. Since $\phi_{\pm 1/2}$ are the unphysical NG bosons, 
    the DM candidates in our model are the vector gauge bosons $\chi_\mu$ and $\bar \chi_\mu$.

 To study the spectra of $SU(2)_X$, we have to determine  the nonvanishing VEVs of $H$ and $\Phi_4$. Using Eqs.~(\ref{eq:vphi4}) and (\ref{eq:Nphi4}), we get 
  \be
  V(v, v_4) = \frac{v^2 \mu^2}{2} + \frac{\lambda v^4}{4} + \frac{\mu^2_\Phi v^2_4}{2} + \frac{\lambda_\Phi v^4_4}{4}
 + \frac{\lambda' v^2 v^2_4}{4} \,.
  \ed
 With minimal conditions $\partial V(v, v_4 )/\partial v=\partial V(v,v_4 )/\partial v_4  =0$, we  have
 \be
  \mu^2 + \lambda v^2 + \frac{\lambda' v^2_4}{2} = 0 \,, \non \\
   \mu^2_\Phi + \lambda_\Phi v^2_4 + \frac{\lambda' v^2 }{2} = 0\,, \label{eq:mini}
 \ed
 respectively.
 In terms of the parameters in the scalar potential, the VEVs could be written as
  \be
v^2 &=& \frac{2\lambda' \mu^2_\Phi- 4 \lambda_\Phi \mu^2 }{4 \lambda \lambda_\Phi - \lambda'^2} \,, \non \\
v^2_4 &=&  \frac{2\lambda' \mu^2 - 4 \lambda \mu^2_\Phi }{4 \lambda \lambda_\Phi - \lambda'^2}   \,.
   \ed
As known that the masses of gauge bosons  arise from the kinetic term of $\Phi_4$, accordingly  the masses of  $\chi_\mu (\bar\chi_\mu)$ and $X^3_\mu$ can be directly found by
  \be
&& \Phi^\dagger_0 g^2_X \left[ (T^- T^+ +T^+ T^- ) \chi_\mu \bar\chi^\mu + (T^3)^2 X^3_\mu X^{3\mu} \right] \Phi_0  \non \\
 && = \frac{g^2_X v^2_4}{2} \left[ 2\Big(t (t+1) -t_3^2  \Big) \chi_\mu \bar\chi^\mu + t_3^2 X^3_\mu X^{3\mu} \right]\,,
  \ed
  where $t (t+1)$ and $t_3$ are the eigenvalues of $T^2=T^a T^a$ and $T^3$, respectively. With $t=t_3=3/2$, the masses of gauge bosons are obtained as
   \be
   m_{\chi} = \frac{\sqrt{3}}{2} g_X v_4\,, \ \ \ m_{X^3} = \frac{3}{2} g_X v_4\,. \label{eq:gmass}
   \ed
   
Although there are four scalar fields in the quadruplet, three of them become the longitudinal polarizations of gauge bosons ($\chi_\mu$, $\bar\chi_\mu$, $X^3_\mu$). Therefore, combining with the Higgs doublet in the SM, the remaining physical scalar bosons in the model  are $\phi$ and $\phi_r$.  In terms of the scalar potential in Eq.~(\ref{eq:vphi4}),   the mass matrix for  $\phi$ and $\phi_r$  is expressed by 
  \be
  M^2   =   \left( \begin{array}{cc}
    m^2_\phi & \lambda' v v_4  \\ 
   \lambda' v v_4 & m^2_{\phi_r} \\ 
  \end{array} \right) \label{eq:Mass}
  \ed
with $m_\phi =\sqrt{2\lambda } v$ and $m_{\phi_r} =\sqrt{2\lambda_\Phi } v_4$. Due to the  $\lambda'$ effect,   the SM Higgs $\phi$ and $\phi_r$ will mix and are not physical eigenstates.  The mixing angle connected with the mass eigenstates could be parametrized by 
 \be
  \left( \begin{array}{c}
    h \\ 
    H^0 \\ 
  \end{array} \right) =  \left( \begin{array}{cc}
    \cos\theta & \sin\theta \\ 
    -\sin\theta & \cos\theta \\ 
  \end{array} \right)  \left( \begin{array}{c}
    \phi \\ 
    \phi_r \\ 
  \end{array} \right) \,, \label{eq:mixing}
 \ed
where $h$ denotes the SM-like Higgs, $H^0$ is the second scalar boson and  $\tan2\theta = 2\lambda' v v_4 /(m^2_{\phi_r} - m^2_\phi)$. According to Eq.~(\ref{eq:Mass}), the mass squares of physical scalars are found by
 \be
 m^2_{1,2} = \frac{1}{2} \left(m^2_\phi + m^2_{\phi_r}  \pm \sqrt{(m^2_\phi - m^2_{\phi_r})^2 + 4 \lambda'^2 v^2 v^2_4} \right)\,.
 \ed
  We note that the mass of $h$ could be $m_1$ or $m_2$ and  the mass assignment depends on the chosen scheme of the parameters. 
  To solve the problem of the gamma-ray excess, we will focus on the case of $m_h > m_{H^0}$. 
 
 Next, we derive the couplings of $\phi$ and $\phi_r$ and the interactions with new gauge bosons. We first discuss the gauge interactions of $\phi_r$. From Eq.~(\ref{eq:lang}), we see that the gauge interactions of the quadruplet only occur in the kinetic term of $\Phi_4$. Using  $\bar \Phi_4$ defined in Eq.~(\ref{eq:Nphi4}) and the covariant derivative of $\Phi_4$, the gauge interactions are expressed as
  \be
   I_{G} &=&\partial_\mu\bar\Phi^\dagger_4 \left( i g T^a X^{a\mu} \right) \bar\Phi_4 + h.c.  \label{eq:IG}\,, \\
 I_{GG} &=&  \left(i g T^a X^a_\mu \bar\Phi_4 \right)^\dagger  \left(i g T^b X^{b \mu} \bar\Phi_4 \right) \label{eq:IGG}\,.
  \ed 
 By adopting the expression of Eq.~(\ref{eq:T+-}), one can easily find that the gauge interactions of Eq.~(\ref{eq:IG}) vanish. 
  By using  the result
  \be
  T^a X^a_\mu \bar\Phi_4 = \left(   \begin{array}{c}
   3/2\, X^3_\mu  \\ 
    \sqrt{3/2}\, \bar\chi_\mu \\ 
   \sqrt{3/2} \, \chi_\mu \\ 
    - 3/2\, X^3_\mu \\ 
  \end{array}
  \right)\frac{v_4 + \phi_r}{2}\,,
  \ed
  Eq.~(\ref{eq:IGG}) can be straightforwardly  written as 
  \be
 I_{GG}& =& \sqrt{3} g_X m_\chi \phi_r \chi_\mu \bar\chi^\mu + \frac{3 \sqrt{3}}{2} g_X m_\chi \phi_r X^3_\mu  X^{3\mu} \non \\
 &+& \frac{1}{2} \left(\frac{3 g^2_X}{2} \right) \phi^2_r \chi_\mu \bar\chi^\mu + \frac{1}{4} \left( \frac{9 g^2_X}{2}\right) \phi^2_r X^3_\mu X^{3\mu}\,, \label{eq:igg}
  \ed
  where  the masses of gauge bosons defined in Eq.~(\ref{eq:gmass}) have been applied. 
We second discuss the couplings of $\phi_r$ to the SM Higgs $\phi$ where the vertices  could be obtained from the scalar potential of Eq.~(\ref{eq:vphi4}).  Since the derivations are straightforward, we summarize the vertices of $\phi_r$ and $\phi$ in Table~\ref{tab:sint}. We note that although  the interactions  in Eq.~(\ref{eq:igg}) and Table~\ref{tab:sint} are shown in terms of $\phi_r$ and $\phi$,   the expressions with $h$ and $H^0$ mass eigenstates  could be easily obtained when Eq.~(\ref{eq:mixing}) is applied. 
  \begin{table}[htbp]
   \caption{Couplings of the scalar boson $\phi_r$ to SM Higgs $\phi$. }
  \label{tab:sint}
  \begin{ruledtabular}
  \begin{tabular}{ccccc} 
   $\phi_r \phi^2 $ & $\phi^2_r  \phi$  & $\phi^3_r $  & $\phi^2_r  \phi^2 $ & $\phi^4_r$   \\  \hline
    $ \lambda' v_4 $ & $\lambda' v$ & $3! \lambda_\Phi v_4$ & $\lambda'$ & $3! \lambda_\Phi$ \\ 
  \end{tabular}
  \end{ruledtabular}
\end{table}
  
 The relevant free parameters in the model are $\mu^2_{(\Phi)}$, $\lambda_{(\Phi)}$, $\lambda'$ and the gauge coupling $g_X$. Using the masses of $\phi$ and $\phi_r$ and the VEVs of $H$ and $\Phi_4$, the six parameters could be replaced by $(g_X, v, v_4, m_\phi, m_{\phi_r}, \lambda')$.  When these values of parameters are fixed, the masses of $h$ and $H^0$ and the mixing angle $\theta$ are  determined. According to the results measured by ATLAS \cite{:2012gk} and CMS \cite{:2012gu}, the Higgs mass now is known to be $m_h=125$ GeV. Therefore, it is better to use the physical masses $m_{h, H^0}$ and mixing angle $\theta$ instead of $m_{\phi,\phi_r}$ and $\lambda'$.  
 Additionally,   the VEV of $ v\approx 246$ GeV is determined  from the Fermi constant $G_F$ and $v_4$ can be replaced by $m_\chi$. Hence, the involving  unknown parameters in the model are $g_X$, $m_{\chi}$, $m_{H^0}$ and $\theta$. 
 
\begin{figure}[hpbt] 
\begin{center}
\subfigure[]{\includegraphics[width=50mm]{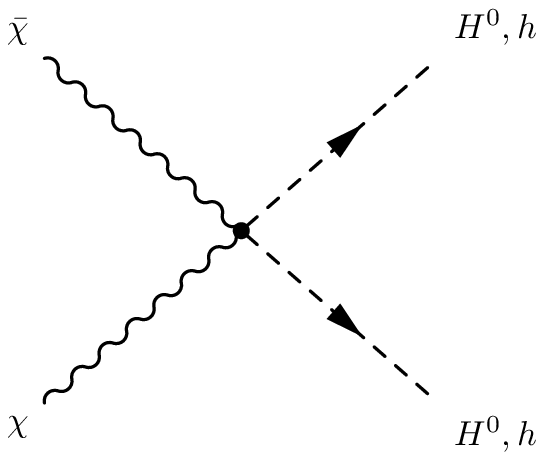} }
\subfigure[]{\includegraphics[width=50mm]{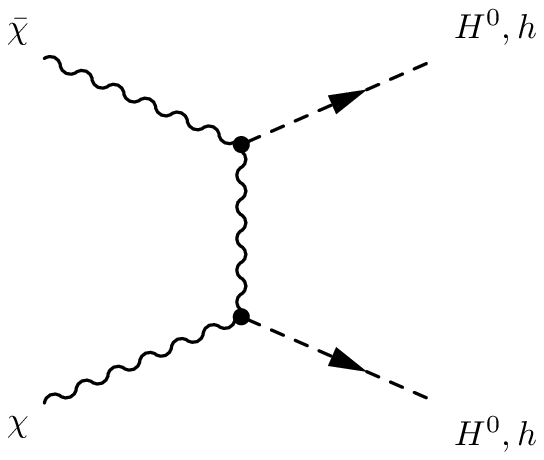}}
\subfigure[]{\includegraphics[width=50mm]{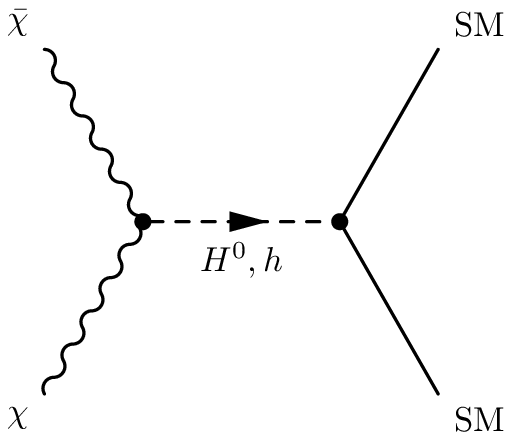}}
\caption{ Processes of the DM annihilation. Diagram II includes $t$- and $u$-channel. }
\label{fig:Feyns}
\end{center}
\end{figure}

To constrain the free parameters, two observables have to be   taken into account:  one is the relic density~\cite{Ade:2013ktc} and another one is 
the DM-nucleon scattering cross section \cite{XENON100,Akerib:2013tjd}. The number density of DM is dictated by the well-known Boltzmann equation, expressed by
 \be
 \frac{dn}{dt} + 3 {\bf H} n = -  \langle \sigma v_{\rm rel} \rangle \left(n^2 - n^2_{\rm eq} \right) \label{eq:Beq}
 \ed
where $\bf H$ is the Hubble parameter,  $n=n_\chi + n_{\bar\chi}$, and $n_{\rm eq}$ is the equilibrium density, defined by   
 \be
 n_{\chi, \rm eq} =n_{\bar\chi, \rm eq}= g_{\chi} \frac{m^2_\chi T}{2\pi^2} K_2 \left( \frac{m_\chi}{T}\right) \,,
   \ed
 with $g_{\chi}$  the internal degrees of freedom of DM, $T$  the temperature and  $K_i$ the modified Bessel function of the second kind \cite{Gondolo:1990dk}. For the vector DM, we take $g_\chi=3$. The thermally averaged annihilation cross section is given by
  \be
  \langle \sigma v_{\rm rel} \rangle = \frac{1}{8T m^4_{\chi}  K^2_2(m_\chi/T)} \int^\infty_{4m^2_\chi } ds \sqrt{s} ( s - 4m^2_\chi) K_1(\sqrt{s}/T)\sigma(\chi \bar\chi \to {\rm all})\,.
  \ed
 In the model, the DM annihilating  into the SM particles  is through the Higgs portal, where  the associated Feynman diagrams  are presented in Fig.~\ref{fig:Feyns}. 
 We note that in contrast to Ref.~\cite{Boehm:2014bia},  the DM semi-annihilation processes such as $\chi \chi \to \chi (H^0,\, h)$ are absent in our model.
 %
%
 To study the DM abundance after the freeze-out, usually it is more convenient to consider the ratio of the number density to entropy density, defined by  $Y=n/s$,  where $s= (2\pi^2/45) g_* T^3$ and $g_*(T)$ is the effective number of degrees of freedom contributing to the entropy density. With ${\bf H} = - \dot{T} /T$, $\dot{s} + 3 H s =0$ and $x=m_\chi/T$, Eq.~(\ref{eq:Beq}) leads to
 \be
\frac{dY}{dx} \approx  -\frac{4\pi}{ \sqrt{90}} \frac{m_\chi M_P}{x^2} \sqrt{g_*(T)} \langle \sigma \rm v_{rel}\rangle \left( Y^2 - Y^2_{eq}\right) \,,\label{eq:BY}
 \ed
where $H^2 = 8\pi^3 G g_* T^4/90$ and  $M^2_P = 1/(8\pi G)$ have been used. If we set $Y_\infty$  to be the present value after the freeze-out, the current  relic density of DM  is given by
 \be
 \Omega_\chi = \frac{m_\chi s_0 Y_\infty}{3 H^2_0 M^2_P}\,, \label{eq:RD}
 \ed
where $H_0$ and $s_0$ are the present Hubble constant and entropy density, respectively.  For numerical calculations, we employ { \tt micrOMEGAs 4.1.5 } \cite{Belanger:2014vza} to solve the Boltzmann equation  and  get the present relic density of DM  defined in Eq.~(\ref{eq:RD}). 

\begin{figure}[hpbt] 
\includegraphics[width=2.7 in]{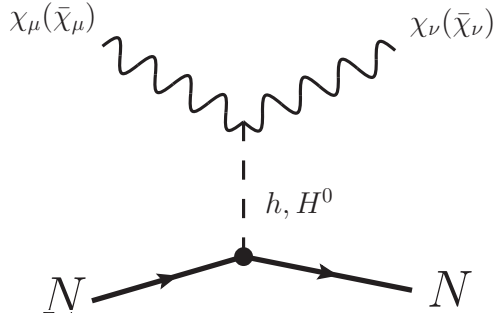} 
\caption{  The sketch of the dark matter scattering off a nucleon.} 
\label{fig:chiN}
\end{figure}

Although the direct detection of DM via the DM-nucleon scattering has not been observed yet, the  sensitivity of the current experiment could give a strict constraint on the free parameters. In the model, the sketch of a vector DM scattering off a nucleon is shown in Fig.~\ref{fig:chiN}. 
By neglecting the small momentum transfer, the scattering amplitude of the $\chi_\mu(\bar\chi_\mu)$-nucleon is written as
\begin{equation}
M = \epsilon^\mu \epsilon^*_\mu(k_1) \frac{\sqrt{3} g_X m_\chi}{v} \frac{m^2_h - m^2_{H^0}}{m^2_h m^2_{H^0}} \sin \theta \cos \theta \langle N | m_q \bar q q |N \rangle\,.
\end{equation}
By assuming that  the effective couplings of DM to the proton and  neutron are the same,  we parametrize the nucleon transition matrix element to be $\langle N | m_q \bar q  q | N \rangle = f_N/(\sqrt{2}G_F)^{1/2}$, where the  range of $f_N$ is  $[1.1, 3.2]\times 10^{-3}$~\cite{Cheng:2012qr, He:2011de}.   As a result, the scattering cross section of the DM-nucleon is  formulated by 
 \be
\sigma_{\chi N} = \sigma_{\chi (\bar\chi) N \to \chi(\bar\chi) N} \approx  \frac{3 g^2_X f^2_N }{4\pi} (\sin\theta \cos\theta)^2\left(\frac{m_N}{m_\chi + m_N}\right)^2  \left( \frac{m^2_{H^0} - m^2_h}{m_h m_{H^0}}\right)^2. \label{eq:DMscattering}
 \ed
 
Before discussing the numerical analysis, we set up the possible schemes for the values of $m_\chi$ and $m_{H^0}$.  Since $\chi \bar \chi \to W^+ W^-, ZZ$ are the dominant channels in the case of  $m_{H^0} > m_\chi > m_W$ and  in disfavor with  the gamma-ray spectrum~\cite{Calore:2014nla}, 
 we assume $\chi$ is lighter than $W$ and $Z$.   To explain the excess of the gamma-ray spectrum, it has been pointed out that the preferred channels via the Higgs portal are $\chi \chi\to S S \to b \bar b b \bar b$ with $S$ being the possible scalar and $\chi \chi \to b \bar b$~\cite{Calore:2014nla,Berlin:2014pya,Boehm:2014bia}, where the former produces the on-shell $S$ and  subsequently  $S$ decays into SM particles while the latter utilizes the resonant enhancement of $m_S \sim 2 m_\chi$. 
As a result,  we focus on the following two schemes:  

(a) $m_\chi = 70\,, 60$ GeV and  $m_{H^0} < m_\chi$, where the DM annihilation channel is  $\chi \bar \chi \to H^0 H^0$ with $H^0$ being 
the on-shell scalar boson; and afterwards $H^0$ decays through $H^0 \to b \bar b$ \cite{Berlin:2014pya,Boehm:2014bia}. 

(b) $m_\chi = 50\,,40$ GeV and $m_{H^0} > m_\chi$, where the DM annihilation channel is $\chi \bar \chi \to b \bar b$~\cite{Calore:2014nla,Boehm:2014bia}. We will see that the channel becomes significant when the condition of $m_{H^0}\sim 2 m_\chi$ is satisfied. 

\noindent 
Although the fermions in the final states could be other lighter leptons and quarks, since the coupling of the scalar to the fermion  depends on the mass of the fermion,  we only focus on the $b$-quark pairs in the final states.  

\begin{figure}[hptb] 
\begin{center}
\includegraphics[width=70mm]{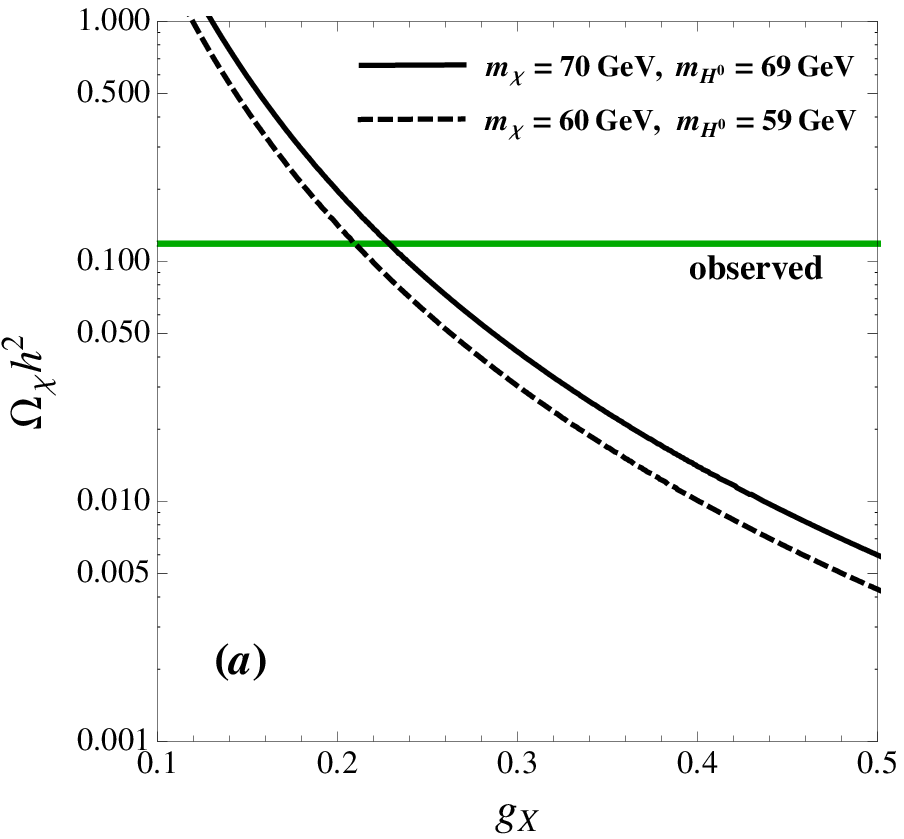} 
\includegraphics[width=70mm]{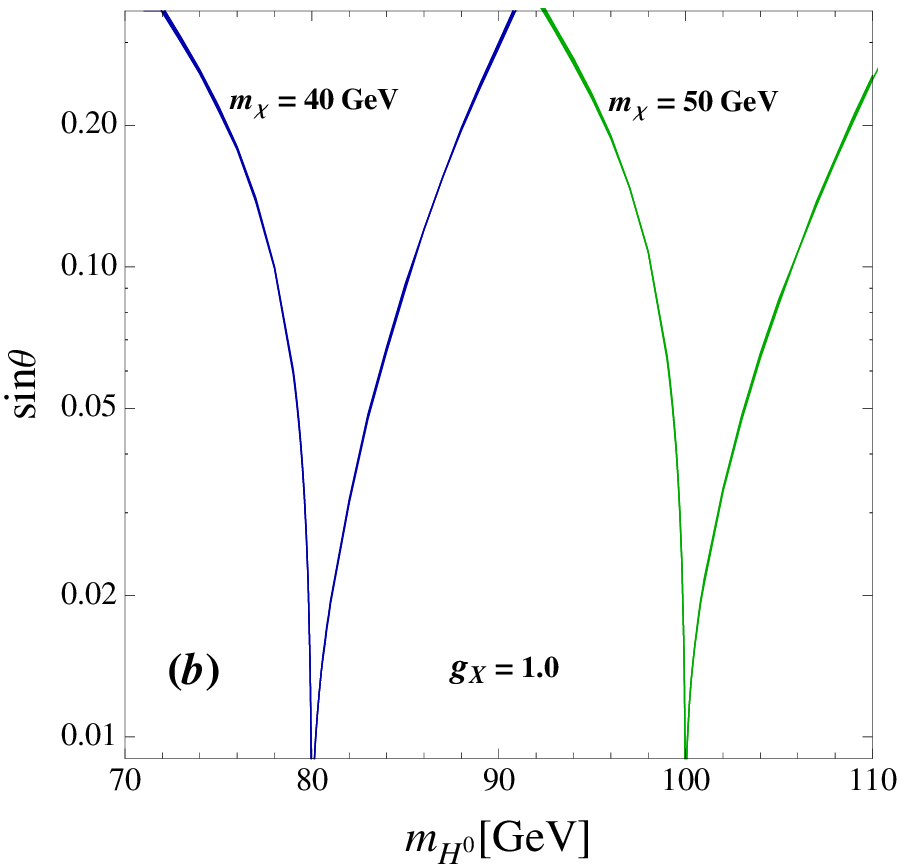} 
\caption{ (a) Relic density of $\chi(\bar \chi)$ as a function of $g_X$ in scheme (a), where  the band indicates the observed value of $\Omega_{\rm DM} h^2$. (b) Correlation between $\sin\theta$ and $m_{H^0}$ in scheme (b) when $g_X=1$ is taken and the observed relic density of DM is satisfied. }
\label{fig:RD1}
\end{center}
\end{figure}

In scheme (a), as the main DM annihilating processes are from Figs.~\ref{fig:Feyns}I and \ref{fig:Feyns}II and the produced $H^0$ pairs are on-shell, 
 the results are insensitive to the mixing angle $\theta$. To understand the constraint of the  observed $\Omega_{\rm DM} h^2$, we present 
 $\Omega_\chi h^2$ as a function of $g_X$ in Fig.~\ref{fig:RD1}(a). From the results, we see that for matching the observed relic density of DM,  
 the value of the gauge coupling $g_X$ should be around $0.23(0.21)$ for $m_\chi = 70(60)$ GeV and $m_{H^0} = 69(59)$ GeV. We note that for explaining the excess of the gamma-ray via the DM annihilation, we adopt $m_\chi \approx m_{H^0}$ in scheme (a). We will clarify this point later. In scheme (b), Fig.~\ref{fig:Feyns}III becomes dominant. Since $h$ and $H^0$ both contribute to the DM annihilation, besides the gauge coupling $g_X$ and $m_{H^0}$, the results are also sensitive to the mixing angle $\theta$. Since there are three free parameters involved in this scheme, in Fig.~\ref{fig:RD1}(b) we show the correlation between $\sin\theta$ and $m_{H^0}$ when   $g_X=1$ is taken and the observed $\Omega_{\rm DM} h^2$ is  simultaneously satisfied.

\begin{figure}[hpbt] 
\begin{center}
\includegraphics[width=70mm]{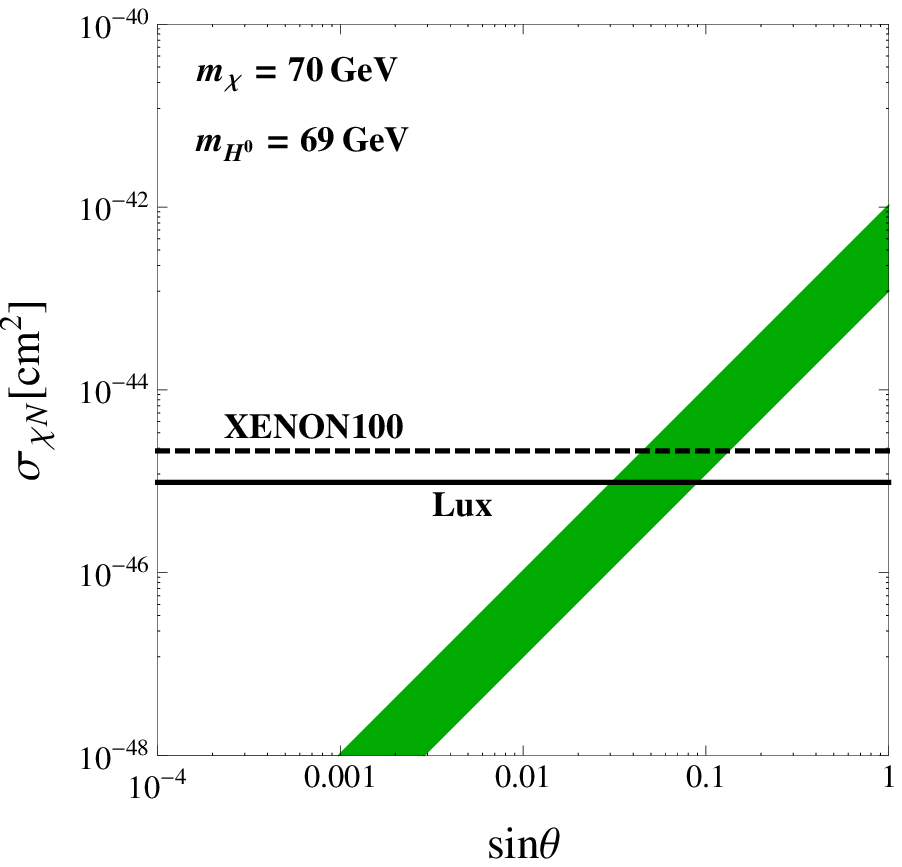} 
\includegraphics[width=70mm]{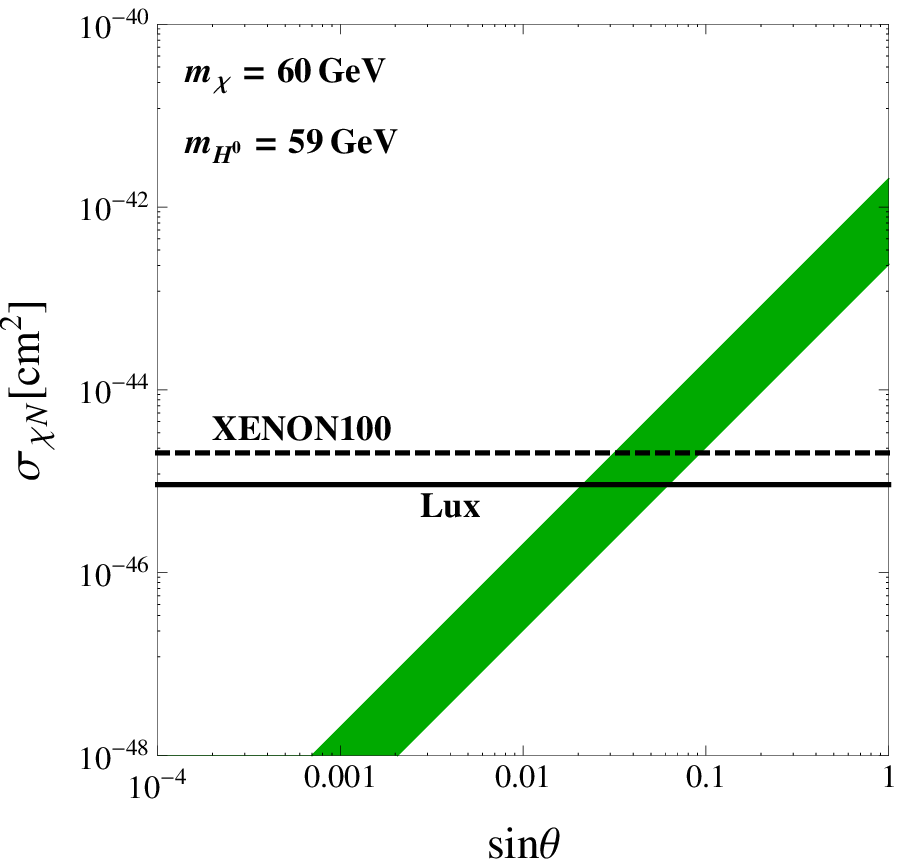} 
\caption{  DM-nucleon scattering cross section in scheme (a), where the constraint of observed $\Omega_{\rm DM} h^2$ has been considered. For comparisons, the measurements of  XENON100~\cite{XENON100} and LUX~\cite{Akerib:2013tjd} for 90\%-CL upper limits are shown in the plots.}
\label{fig:directD1}
\end{center}
\end{figure}

\begin{figure}[hpbt] 
\begin{center}
\includegraphics[width=70mm]{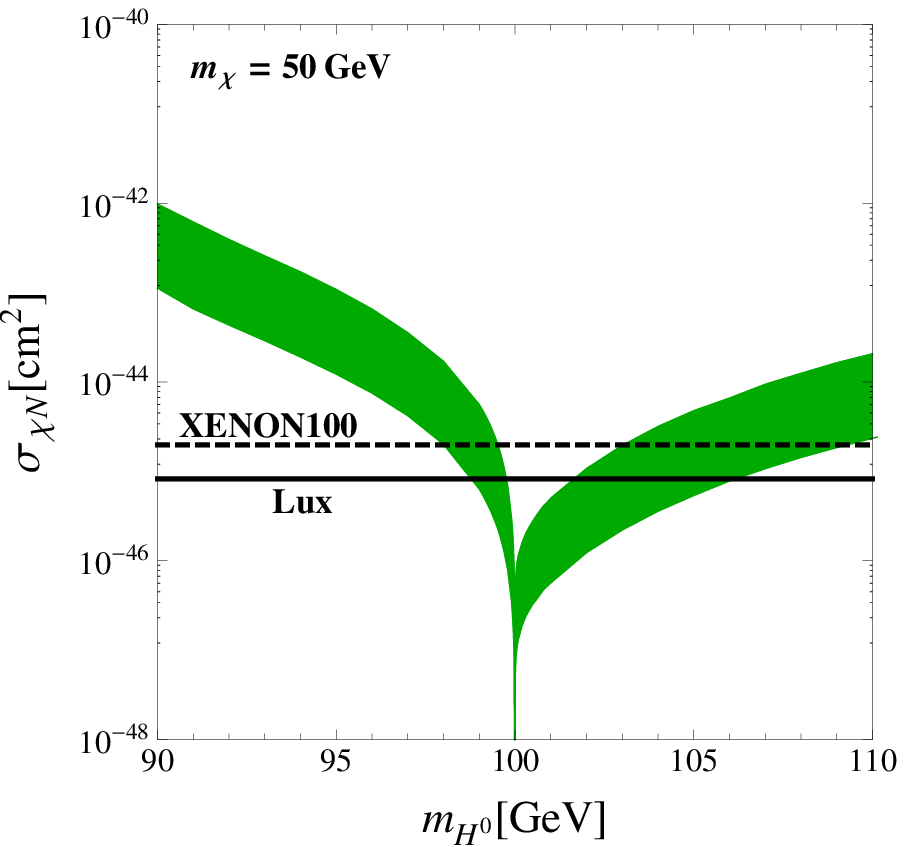} 
\includegraphics[width=70mm]{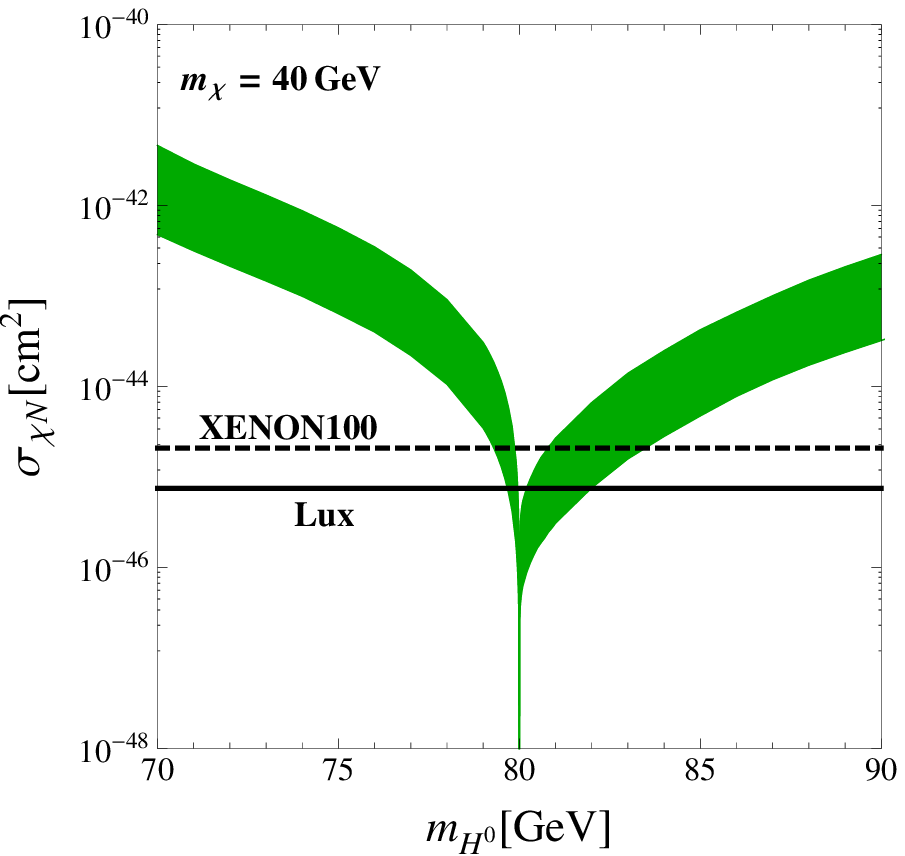} 
\caption{DM-nucleon scattering cross section as a function of $m_{H^0}$ in scheme (b), where the constraint of observed $\Omega_{\rm DM} h^2$ has been considered.  The measurements of  XENON100~\cite{XENON100} and LUX~\cite{Akerib:2013tjd} for 90\%-CL upper limits at the corresponding $m_\chi$ are also shown in the plots. }
\label{fig:directD2}
\end{center}
\end{figure}

%
With the proposed schemes (a) and (b), we can further discuss the constraints 
 from the measurements of DM direct detections. Since the vector DM candidates are not self-charge-conjugation particles, the DM density is composed of $\chi_\mu$ and $\bar\chi_\mu$, i.e. $\rho_{\rm DM} = \rho_\chi + \rho_{\bar \chi}$. Thus, the elastic scattering cross section of DM  off a nucleon is proportional to $ \rho_\chi \sigma_{\chi N} + \rho_{\bar \chi}  \sigma_{\bar \chi N} =  \rho_{\rm DM} \sigma_{\chi N}$. Consequently, for comparing with the DM-nucleon scattering cross section measured by the direct detection experiments, one  can just  use $\sigma_{\chi N}$  which is formulated in Eq.~(\ref{eq:DMscattering}). For scheme (a), unlike $\Omega_\chi h^2$, $\sigma_{\chi N}$ is $\sin\theta$ dependent. We plot the elastic cross section as a function of $\sin\theta$ in Fig.~\ref{fig:directD1}, where we have taken  $(m_\chi, m_{H^0})= (70, 69)$ GeV for the left panel and $(60, 59)$ GeV for the right panel. In order to  fit the measurement of $\Omega_{\rm DM} h^2$ simultaneously, we use $g_X=0.23 (0.21)$ for the former (latter). For comparisons, we also show the 90\%-CL upper limits by  XENON100~\cite{XENON100} and LUX~\cite{Akerib:2013tjd} Collaborations
 on the plots. From the results, we clearly see that  to satisfy the DM direct detection experiments, we need $\sin \theta < 0.1$.   For scheme (b), we present  $\sigma_{\chi N}$ as a function of $m_{H^0}$ in Fig.~\ref{fig:directD2} with $g_X=1$ and $m_\chi= 50~(40)$ GeV for the left (right) panel. In order to fit the data of $\Omega_{\rm DM} h^2$ together, in the figure we have applied the results shown in Fig.~\ref{fig:RD1}(b). By the plots, we find that  current DM direct detection experiments further limit the mass relation to be $m_{H^0} \sim 2 m_\chi$. 

After analyzing the constraints of the DM relic density and direct detection, we now study the gamma-ray which is originated from the DM annihilation. It is known that the flux of the gamma-ray from the DM annihilation is expressed by
\begin{equation}
\frac{d\Phi(E_\gamma, \psi) }{dE_\gamma d \Omega}= \frac{\langle \sigma v_{\rm rel} \rangle}{8 \pi m_\chi^2} \frac{d N_\gamma}{d E_\gamma} \int_{\rm los} \rho^2 (r) dl(\psi)\,, \label{eq:Gr}
\end{equation}
where $dN_\gamma/d E_\gamma$ is the gamma-ray spectrum produced per annihilation, $\psi$ is the observation angle between the line-of-sight and the galactic center, 
$\rho(r)$ is density of DM, and the integration of the density squared is carried out over the line-of-sight.
The general DM halo profile could be parametrized by
 \be
 \rho(r) = \rho_{\odot} \left( \frac{r_\odot}{r} \right)^{\gamma} \left( \frac{1+ (r_\odot/r_s)^\alpha}{1+ (r/r_s)^\alpha} \right)^{(\beta-\gamma)/\alpha}\,,
 \ed
 where $r_s =20$ kpc is the scale radius, $\rho_\odot =0.3$ GeV/cm$^3$ is the local dark matter density at $r_\odot = 8.5$ kpc and $r$ is the distance from the center of the galaxy. Note that $(\alpha, \beta,\gamma)=(1, 3, 1)$ corresponds to the Navarro-Frenk-White (NFW) profile. In our numerical estimations, we set $\alpha=1$ and $\beta=3$, but $\gamma$ to be a free parameter.  
 Since $\rho(r)$ is proportional to $r^{-\gamma}$, we see that the change of the parameter $\gamma$  can  only shift the entire gamma-ray spectrum  but not the shape of gamma-ray flux. 
 For executing the numerical calculations of Eq.~(\ref{eq:Gr}), we implement our model to { \tt micrOMEGAs 4.1.5 }~\cite{Belanger:2014vza}  and use the program code to estimate the gamma-ray spectrum. 
 
\begin{figure}[hpbt] 
\begin{center}
\includegraphics[width=70mm]{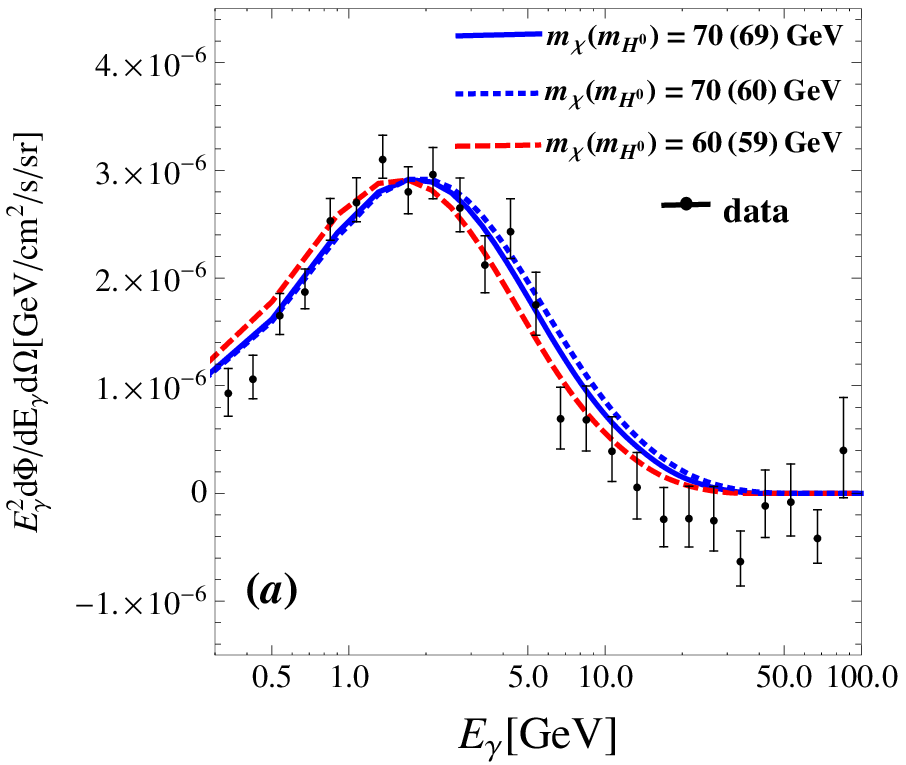} 
\includegraphics[width=70mm]{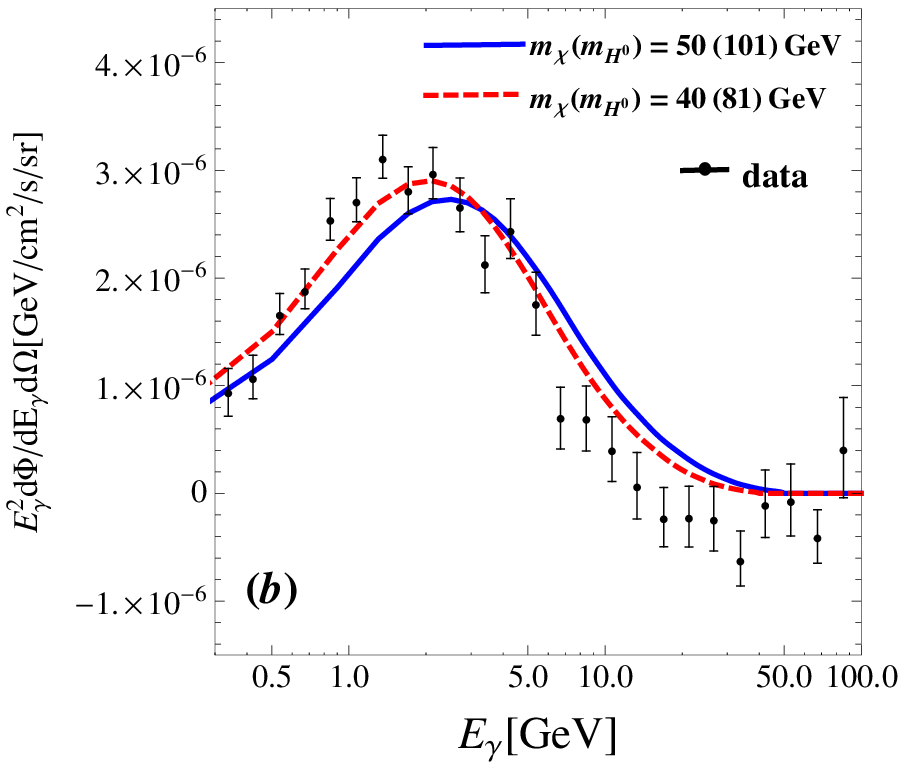} 
\caption{Gamma-ray spectrum from dark matter annihilation processes (a) $\chi \bar \chi \to H^0 H^0$ in which $H^0$ mainly decays into $b \bar b$  and (b) $\chi \bar \chi \to b \bar b$, where the former corresponds to scheme (a) and the latter is scheme (b). The values of slope index $\gamma$ are taken as 1.26 [1.22] for $( m_\chi, m_{H^0} ) = (70, 69 )[ (70, 60 )\  \textrm{and}\  (60, 59)]$ GeV and 1.33 for $m_\chi =50(40)$ GeV. The data are quoted from  Ref.~\cite{Daylan:2014rsa} with $\psi = 5$ degrees .}
\label{fig:GR1}
\end{center}
\end{figure}

In the model, the processes to produce the gamma-ray by the DM annihilation are similar to those for the relic density, except that
the gamma-ray is emitted in the final states.  In scheme (a), we present the flux of the gamma-ray as a function of the photon energy $E_\gamma$ in Fig.~\ref{fig:GR1}(a), where the solid line denotes  $(m_\chi, m_{H^0})=(70,69)$ GeV and $(\gamma, g_X)=(1.26, 0.23)$, the dotted line represents $(m_\chi, m_{H^0})=(70,60)$ GeV and $(\gamma, g_X)=(1.22, 0.21)$, and  the dashed line is $(m_\chi, m_{H^0})=(60,59)$ GeV and $(\gamma,g_X)=(1.23, 0.21)$. The taken values of 
the gauge coupling $g_X$ are determined from the observed DM relic density. From the figure, we see that when the mass difference $m_\chi -m_{H^0}$ becomes larger, due to the boosted $H^0$,  the flux after the peak of the excess tends to be enhanced and disfavors with the data. Hence, we only focus on $m_\chi \approx m_{H^0}$.  In scheme (b), $\chi \bar \chi \to b \bar b$ is dominant. The result of the gamma-ray flux as a function of $E_\gamma$ is given in Fig.~\ref{fig:GR1}(b), where $g_X=1$ is taken,  the solid and dashed lines stand for $(m_\chi, m_{H^0})= (50, 101)$ and $(40, 101)$ GeV, respectively, and  the value of $\sin\theta \simeq 0.02$ is read from  Fig.~\ref{fig:RD1}(b) for both cases when the observed $\Omega_{\rm DM} h^2$ is satisfied.  In addition, the value of $m_{H^0}$ has been chosen to follow the constraint of the direct detection, i.e. $m_{H^0} \sim 2 m_{\chi}$.  For the case of $m_{H^0} \lesssim 2 m_\chi$, due to the produced $H^0$ being an on-shell particle,  the annihilation cross section becomes too large to explain the gamma-ray excess. Hence, we adopt $m_{H^0} \gtrsim 2 m_\chi$. 

 Finally, we make some comparisons with the study in  Ref.~\cite{Boehm:2014bia} where the stable DM candidates are dictated by the custodial symmetry~\cite{Hambye:2008bq}. Since the trilinear couplings of gauge bosons exist in the model given by Ref.~\cite{Hambye:2008bq}, besides the annihilation processes which we only have in our model, there are also semi-annihilation processes in Refs.~\cite{Boehm:2014bia,Hambye:2008bq}. With the taken values of parameters and the best-fit approach, the authors of Ref.~\cite{Boehm:2014bia} have found that the gamma-ray excess is dominated by the semi-annihilation. As a result, DM with its mass around $39-76$ GeV could fit the measured gamma-ray spectrum of the Galactic Center. However, the resulted $\langle \sigma v_{\rm rel} \rangle$ is  a factor of 2-3 larger than  that of the observed $\Omega_{\rm DM} h^2$. In our approach, with the selected values of $m_\chi$, e.g. $m_{\chi}=(70,60)$ GeV in scheme (a) and $m_\chi=(50, 40)$ GeV in scheme (b), we first constrain the free parameters  by using the observed  $\Omega_{\rm DM} h$ and  the upper limit of the DM direct detection. With the allowed values of parameters, we subsequently  estimate the gamma-ray spectrum from the DM annihilation. Although the best-fit approach is not adopted in the analysis,  our results from the on-shell $H^0$ production in scheme (a) and $m_{H^0} \sim 2 m_{\chi}$ in scheme (b) are morphologically consistent with the  gamma-ray spectrum  of 
 the Galactic Center.


 In summary, to interpret the excess of the gamma-ray through the DM annihilation, we have studied the DM model in the framework of  $SU(2)_X$ gauge symmetry. To break the gauge symmetry, we have used one quadruplet of $SU(2)_{X}$. As a result, the remnant $Z_3$ symmetry of $SU(2)_X$ leads to the stable DMs, which  are the  gauge bosons of $SU(2)_X$. Due to the mixture of the quadruplet and SM Higgs doublet in the scalar potential, the DM annihilation to SM particles is through the Higgs portal. When the observed relic density of DM and the limit of the DM direct detection are both satisfied, we find that  $m_\chi < m_{W}$ could give a correct pattern for the gamma-ray spectrum. For more specific numerical studies, we classify the values of parameters to be scheme (a) with $(m_\chi, m_{H^0})=(70, 69)$ and $(60, 59)$ GeV and scheme (b) with $(m_\chi , m_{H^0})=(50, 101)$ and $(40, 81)$ GeV. We show that for matching the  gamma-ray excess, in scheme (a) it is better to take $m_\chi \approx m_{H^0}$. If $m_\chi - m_{H^0}$  is increasing, due to the boosted $H^0$, the gamma-ray flux  at the photon energy  over the peak of the gamma-ray spectrum  is  enhanced and the resulted flux tends to be away from the data. In scheme (b), for avoiding the constraint from the DM direct detection and  the  production of the on-shell $H^0$ which causes too large cross section,  the condition of $m_{H^0} \gtrsim 2m_\chi$ is adopted.  Based on our current analysis,  we see that the results of scheme (a) fit the data well.  \\

\noindent{\bf Acknowledgments}

 This work is supported by the Ministry of Science and Technology of 
R.O.C. under Grant \#: MOST-103-2112-M-006-004-MY3 (CHC) and MOST-103-2811-M-006-030 (TN). We also thank the National Center for Theoretical Sciences (NCTS) for supporting the useful facilities. 

\end{document}